\documentclass{elsart}
\journal{New Astronomy}

\usepackage{natbib}

\usepackage{graphicx}
\usepackage{epsfig}

\usepackage{amssymb}

\begin{document}

\begin{frontmatter}



\title{HI 21cm probes of reionization, and beyond}


\author{C.L. Carilli$^1$}
\address{$^1$National Radio Astronomy Observatory, Socorro, NM, USA, 87801\\
E-mail: ccarilli@nrao.edu}

\begin{abstract}

I review the potential for observing cosmic reionization using the HI
21cm line of neutral hydrogren. Studies include observations of the
evolution of large scale structure of the IGM (density, excitation
temperature, and neutral fraction), through HI 21cm emission, as well
as observations of small to intermediate scale structure through
absorption toward the first discrete radio sources.  I summarize
predictions for the HI signals, then consider capabilities of
facilities being built, or planned, to detect these signals. I also
discuss the significant observational challenges.

\end{abstract}

\begin{keyword}
cosmology \sep radio \sep lines 

\end{keyword}

\end{frontmatter}


\section{Introduction}

The 21cm line of neutral hydrogren presents a unique probe of the
evolution of the neutral intergalactic medium (IGM), and cosmic
reionization.  Furlanetto \& Briggs \cite{FB04} point out some of the
advantages of using the HI line in this regard: (i) unlike Ly$\alpha$
(ie. the Gunn-Peterson effect), the 21cm line does not saturate, and
the IGM remains 'translucent' at large neutral fractions \cite{CGO02}.
And (ii) unlike CMB polarization studies, the HI line provides full
three dimensional (3D) information on the evolution of cosmic structure,
and the technique involves imaging the neutral IGM directly, and hence
can easily discern different reionization models \cite{BL04,
FSH03}. HI 21cm observations can be used to study evolution of cosmic
structure from the linear regime at high redshift (ie. density-only
evolution), through the non-linear, 'messy astrophysics' regime
associated with luminous source formation.  As such, HI measurements
are sensitive to structures ranging from very large scales down to the
source scale set by the cosmological Jeans mass, thereby "making 21cm
the richest of all cosmological data sets" \cite{BL04}.

Early calculations of HI 21cm signal from a neutral IGM discuss a
broad range of models for large scale HI structure, including
Zeldovich 'pancakes' arising in HDM models, explosive structure
formation models, and CDM \cite{SZ72, HR79, SR90}.  Early
(unsuccessful) searches for large scale structure in HI based on this
wide range of predictions include searches for pancakes with masses $>
10^{14}$ M$_\odot$ at $z \sim 3$ \cite{uson, oort84, davies78,
debruyn88}, and for even more massive structures ($10^{15}$ M$_\odot$)
at $z = 8.4$ \cite{bebbington86}. Fortunately, the cosmological
parameter space has been greatly reduced due to the advent of the
concordance model of $\Lambda$CDM structure formation, and this paper
will assume the standard precision cosmology parameters
\cite{spergel03}.

In this review I will summarize the HI 21cm probes of cosmic
reionization and the neutral IGM. I will focus on the latest
predictions for the expected signals, and the observational
capabilities of telescopes being built, or planned, to detect these
signals. I begin with a short basic review of the physical processes
involved in reionization that relate directly to the HI 21cm signal.
Previous reviews that dealt with, at least in part, the HI 21cm
signal from cosmic reionization, include \cite{BL00, LB00, 
meiksin99}, and in the context of the Square Kilometer Array (SKA)
\cite{Carilli04a, FB04}. See also articles by Furlanetto, Morales, 
and Peterson in this volume. 

\section{The physics of the neutral IGM}

The physics and equations of radiative transfer of the HI 21cm line
through the neutral IGM have been considered in detail by many authors
\cite{SR90, BA04, FB04, ZFL04, SCK05, tozzi99, MMR97, Morales04}, and
I review only the basic results here.

In analogy to the Gunn-Peterson effect for Ly$\alpha$ absorption
by the neutral IGM, the optical depth, $\tau$, of the neutral 
hydrogen to 21cm absorption for our adopted values of 
the cosmological parameters is:
\begin{equation}
\tau 
 = {{3c^3 h_p A_{10} n_{HI}}\over{32 \pi k_B \nu_{21}^2 T_S H(z)}} \\
  \sim 0.0074 \frac{x_{HI}}{T_S} (1+\delta)(1+z)^{3/2},
  \label{tauofz}
\end{equation}
\noindent This equation shows immediately the rich physics involved in
studying the HI 21cm line during reionization, with $\tau$ depending
on the evolution of cosmic over-densities, $\delta$ (predominantly in
the linear regime), the neutral fraction, $x_{\rm HI}$
(ie. reionization), and the HI excitation, or spin, temperature,
$T_S$.

In the Raleigh-Jeans limit, the observed 
brightness temperature (relative to the CMB) 
due to the HI 21cm line at a frequency 
$\nu = \nu_0/(1+z)$, where $\nu_0=1420.40575$ MHz, is given by: 
\begin{equation}
T_B  \approx ~  \frac{T_S - T_{\rm CMB}}{1+z} \, \tau
\label{eq:dtb} \\
 \approx ~  7 (1+\delta) x_{HI} (1 - \frac{T_{CMB}}{T_S})
(1+z)^{1/2} ~ \rm{mK},
\end{equation}
\noindent The conversion factor from brightness temperature to 
specific intensity, I$_\nu$, is given by: 
$I_\nu = \frac{2 k_B}{(\lambda_{21} (1+z))^2} T_B$, or in more common units:
$I_\nu = 7.1\times 10^4 (1+z)^{-2} T_B$ Jy sr$^{-1}$.
Equation 2 shows that for
$T_S \sim T_{CMB}$ one expects no 21cm signal.
When $T_S >> T_{CMB}$, the brightness temperature
becomes independent of spin temperature. When
$T_S << T_{CMB}$, we expect a strong negative 
(ie. absorption) signal against the CMB. 

An important point to keep in mind is that, for the 21cm experiments
being considered, the signal being observed corresponds to large scale
structure, not individual galaxies. This point is demonstrated in Fig
1, which shows the expected signal in Jy versus redshift for different
HI masses, along with the expected sensitivity of current and future
radio telescopes.  For simplicity, I have assumed a single line width
of 300 km s$^{-1}$, which would correspond to individual virialized
galaxies, or to large scale structures just separating from the Hubble
flow. Fig 1 shows that current instruments, such as the GMRT, can
detect large galaxies ($\sim 10^{10}$ M$_\odot$ in gas) out to only
modest redshifts ($z\le 0.3$), even in long integration times.  At
redshifts corresponding to reionization ($z > 6$), future large area
low frequency radio telescopes, such as LOFAR, and eventually the SKA,
will still be limited to studying large scale structure (HI masses $>
10^{11}$ to 10$^{12}$ M$_\odot$).  Fortunately, the entire IGM is made
up of neutral hydrogen prior to reionization, and the large scale
structure detected is not just density enhancements
(ie. protoclusters), but also structure induced by reionization itself
(the 'bubble machine' of HII regions), and possibly spatial variations
in the spin temperature.

\begin{figure}[!t]
\centerline{\psfig{file=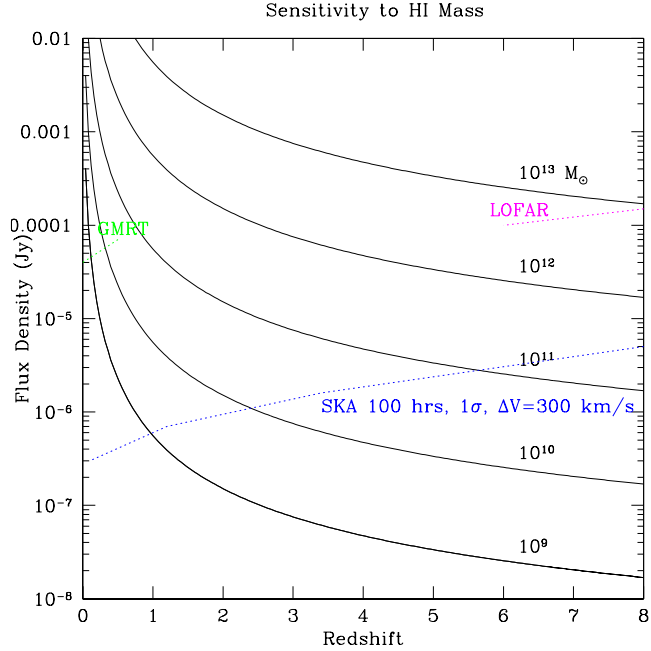,width=3.5in}}
\caption{The HI mass sensitivity versus redshift in 100hrs, for
300 km s$^{-1}$ line width.}
\end{figure}

\cite{tozzi99} show that the HI excitation temperature
will equilibrate with the CMB on a timescale $\sim 
\frac{3\times10^5}{(1+z)}$ year, in absence of other effects. 
However, collisions and resonant scattering of
Ly$\alpha$ photons can drive $T_S$ to the gas kinetic temperature, $T_K$
\cite{field59, wout}:
\begin{equation}
T_S = \frac{T_{\rm CMB} + y_c T_K + 
y_{{\rm Ly}\alpha} T_K}{1 + y_c +  y_{{\rm Ly}\alpha}}. 
\label{eq:hItspin}
\end{equation}
\noindent In this equation, $y_c$ represents collisional excitation of
the hyperfine transition, which couples $T_S$ to the gas kinetic
temperature $T_K$.  The coupling coefficient $y_c \propto n_H$,
and \cite{zygelman05} shows that, for the mean IGM density,
collisional coupling between $T_S$ and $T_K$ becomes 
significant for $z \ge 30$.  The third term in equ 3 corresponds to the
Wouthuysen-Field effect, in which resonant scattering of Ly$\alpha$
photons couples the spin temperature to $T_K$ \cite{wout,field59,
CM03, hirata04, MMR97, meiksin99, PF05}.  \cite{MMR97} show that this latter
mechanism will be important when the radiation background at the
Ly$\alpha$ frequency satisfies $J_\alpha > 9\times 10^{-23} (1+z)$ 
erg cm$^{-2}$ s$^{-1}$ Hz$^{-1}$ sr$^{-1}$, or 
about one Ly$\alpha$ photon per every two baryons at 
$z=8$ \cite{CM03}.

The interplay between the CMB temperature, the kinetic temperature, and
the spin temperature, coupled with radiative transfer,
lead to a number of interesting physical regimes for the HI 21cm
signal. \cite{Ali05, BL04} suggest the following plausible regimes: 

At $z> 200$ residual free electrons couple $T_{CMB}$ and $T_K$ through
Thompson scattering and subsequent gas collisions, while the density
is high enough to equilibrate $T_K$ and $T_S$.  In this case $T_S =
T_{CMB}$ and there is no 21cm signal.

At $z \sim 30$ to 200, the residual ionization fraction and
density is too low to couple $T_K$ to $T_{CMB}$, so the gas cools
adiabatically, with temperature falling as $(1+z)^2$, ie. faster than
the (1+z) for the CMB. Hence the gas becomes colder than the
CMB. However, the mean density in this redshift range is still high 
enough to provide some coupling between  
$T_S$ and $T_K$ through collisions, and the HI 21cm signal
might be seen in absorption against the CMB.  In this regime the HI
fluctuations are still evolving linearly, essentially following the
dark matter \cite{BL04, LZ04, sethi05}.

At $z \sim 20$ to 30, the situation starts to become complex.
Collisions can no longer couple $T_K$ to $T_S$, and $T_S$ again
approaches $T_{CMB}$. However, we might also expect the first luminous
structures (Pop III stars or mini-quasars), at least near the end of
this redshift range.  The Ly$\alpha$ photons from these objects would
induce local coupling of $T_K$ and $T_S$, thereby leading to some 21cm
absorption regions.  On the other hand, \cite{BL04, MMR97} point out
that these same photons, and more importantly, any Xrays from the
first luminous sources \cite{ChenME03}, could lead to local IGM
warming above $T_{CMB}$ well before reionization. Energetically it
takes 13.6 eV per baryon to ionize the IGM, but only $0.005
\frac{(1+z)}{20}$ eV to warm the IGM about $T_{CMB}$. Hence one might
expect a 'patch work' of regions with no signal, absorption, and
perhaps emission, in the 21cm line.

At $z \sim 6$ to 20 all the physical processes come to play. The IGM
is being warmed by the resonant scattering of Ly$\alpha$ photons
\cite{MMR97}, and penetrating, ionizing hard Xrays, from the first
galaxies and black holes \cite{LZ04, BL04, CM03}, as well as by weak
shocks associated with structure formation \cite{FSH03, GS03, SCK05,
ChenME03}, such that $T_K$ is likely larger than $T_{CMB}$ globally
\cite{FSH03}.  Likewise, these objects are reionizing the universe,
leading to a fundamental topological change in the IGM, from the
linear evolution of large scale structure, to a bubble dominated era
of HII regions \cite{FMH05}. It is this regime that most (although not
all), theoretical work has been exploring, and the expected HI 21cm
signal is rich. Finally, after reionization, $z < 6$ to 10, the IGM is
fully ionized ($x_{HI} < 10^{-4}$), and the 21cm signal is gone.

\section{HI 21cm probes of the evolution of the IGM}

\subsection{Global HI signature}

In an idealized sense, if the universe reionized very rapidly
everywhere, at the same time, one would expect a global (ie. full sky)
step in the background temperature at the frequency corresponding to
the redshifted 21cm line \cite{Shaver99}.  However, if nothing else,
cosmic variance will lead to different ionization redshifts in
different regions, and we now know that reionization is likely an
extended process in time, thereby smoothing out the expected signal in
space and time. On the other hand, since this is an all sky signal, the
sensitivity of the experiment is independent of telescope collecting
area, and the experiment can be done using small area telescopes at
low frequency, with well controlled frequency response.

\begin{figure}[!t]
\psfig{file=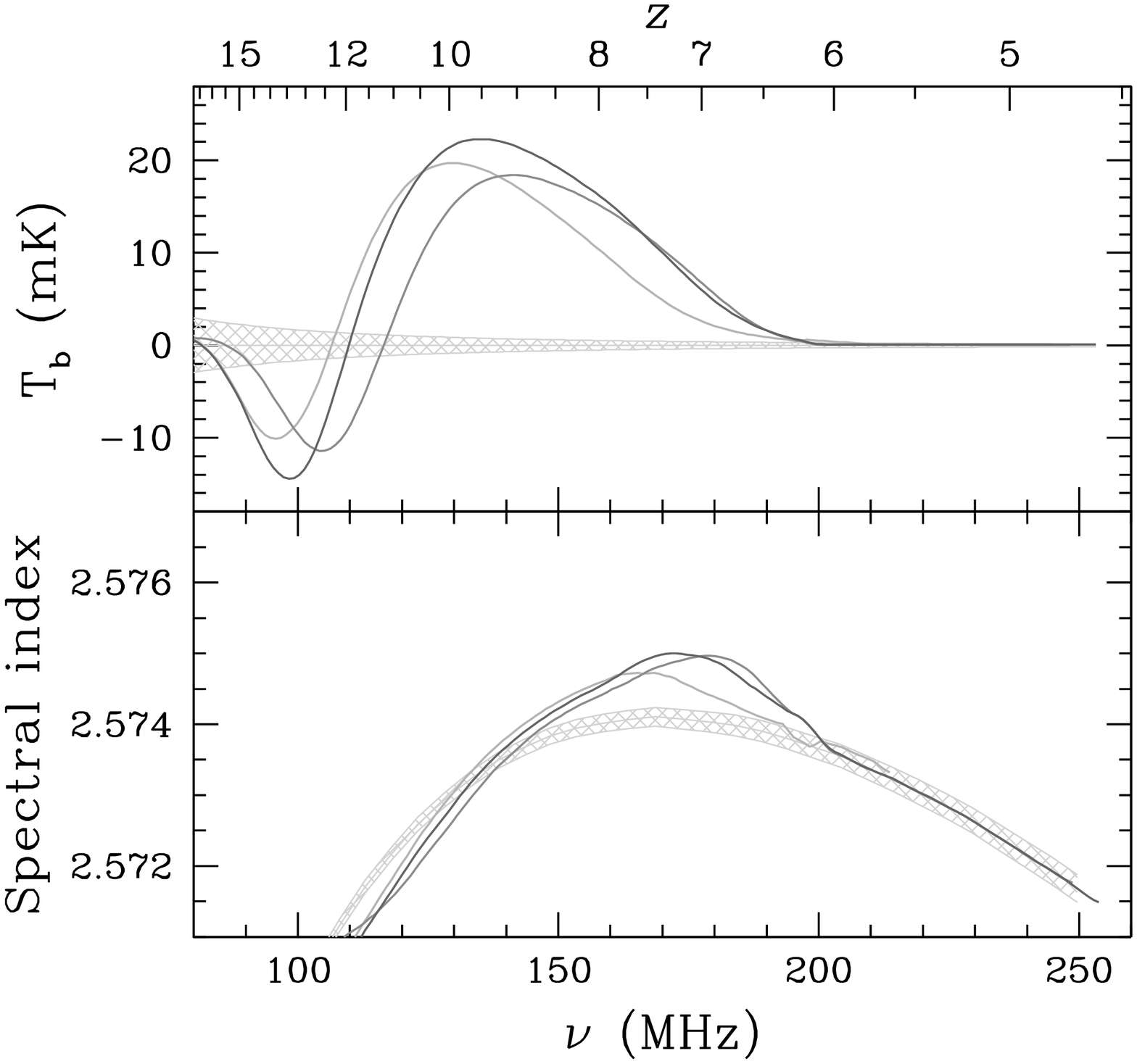,width=2.8in}
\vskip -2.65in
\hspace*{2.85in}
\psfig{file=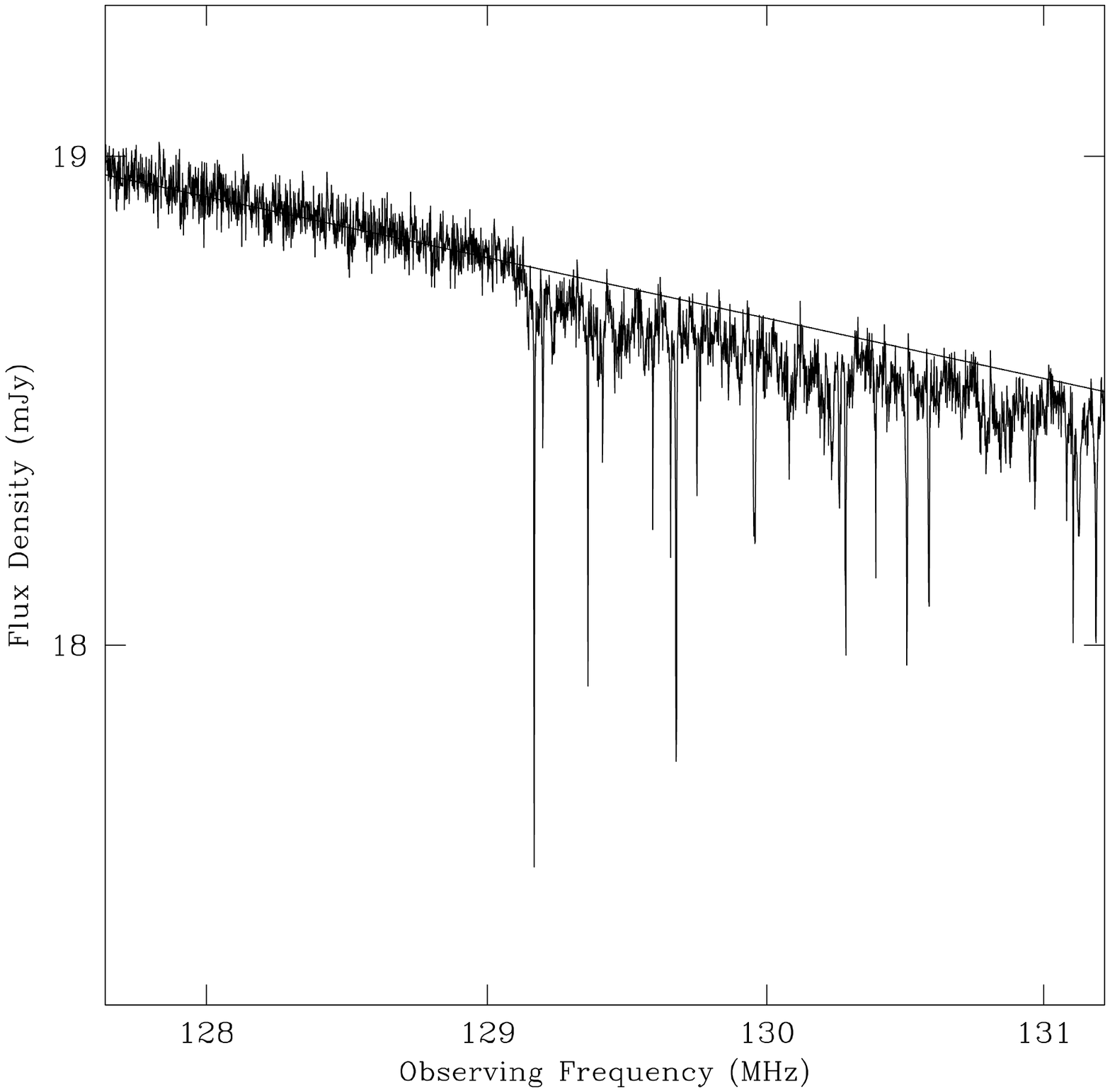,width=2.5in}
\caption{Left: Global (all sky) HI signal from reionization 
\cite{GS03}. The shaded region shows the expected thermal noise
in a carefully controlled experiment. Right: 
The simulated spectrum of a source with S$_{120}$ = 20 mJy at
$z = 10$ using a spectral model based on the powerful radio galaxy
Cygnus A and assuming HI 21cm absorption by the IGM \cite{CGO02}.
Thermal noise has been
added using the specifications of the SKA and assuming 10 days
integration with 1 kHz wide spectral channels.  The onset of
absorption by the neutral IGM is seen at 129 MHz, corresponding to the
HI 21cm line at $z=10$.}
\end{figure}

The most recent modeling of the expected global HI signal is presented
in \cite{GS03} (see also \cite{FSH03}), who employ the numerical
simulations of \cite{Gnedin00}. They include both a fast reionization
model, plus more complicated, {\sl ad hoc} reionization histories.

Results from these calculations are reproduced in Fig 2a.  The
expected signal peaks at roughly 20 mK above the foreground. In their
fiducial (fast reionization) model, the peak occurs at $z \sim 10$.
They also predict a negative signal, meaning absorption against the
CMB, at higher redshift, prior to IGM warming, but allowing for
Ly$\alpha$ resonant scattering (ie. the era of 'Ly$\alpha$ coupling
\cite{BL04}). The shaded region shows the potential system thermal
noise of a well calibrated low frequency experiment.

\cite{GS03} point out that detecting this signal against the mean
non-thermal foreground radiation may be difficult. The foreground is
the sum of relatively smooth Galactic synchrotron emission and
discrete distant radio galaxies, with a typical spectral index, $\alpha
\sim -0.8$, where $\alpha$ is defined as: $S_\nu \propto \nu^\alpha$.
The foreground temperature behaves roughly as $T_{FG} \sim 100
(\frac{\nu}{200 \rm{MHz}})^{-2.8}$, in the coldest regions of the sky,
and can be an order of magnitude higher in the Galactic plane. Hence,
the expect change in $T_B$ is at most $\sim 10^{-4}$ that of the mean
foreground signal over the frequency range $\sim 100$ to 200 MHz.

\subsection{Large scale structure}

\subsubsection{Power spectra and tomography}

The majority of theoretical studies have focused on predicting the HI
fluctuations in the IGM on scales of arcmins to a few degrees both
during \cite{ZS04, MMR97, tozzi99, FZH04, FSH03, BL04, Ali05, BA04,
GS03, SCK05, FMH05, Wang05}, and prior to \cite{CM03, BL05, LZ04},
reionization.  Most studies have focused on the diffuse IGM, although
a few studies have also considered the signal expected due to the
clustering of collapsed structures \cite{Iliev03}.  

\begin{figure}[!t]
\centerline{\psfig{file=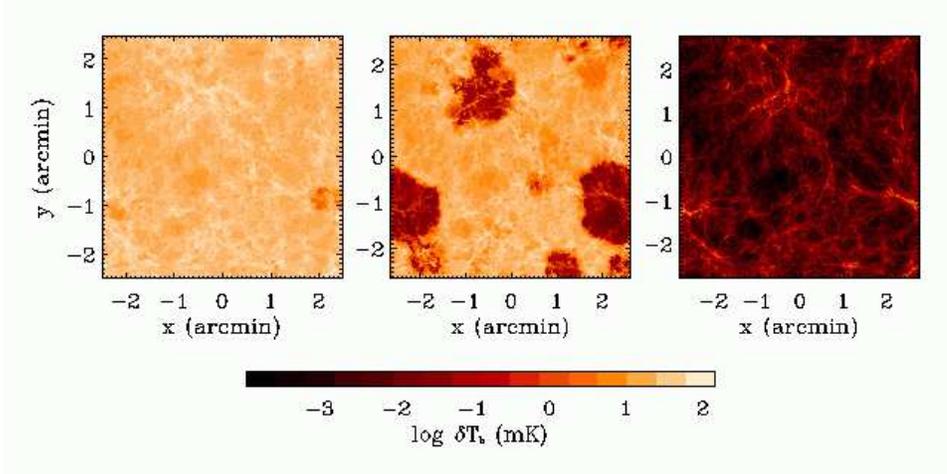,width=5in}}
\caption{The simulated HI 21cm brightness distribution 
during reionization at $z = 12$, 9, 7 \cite{ZFH04, FZH04}.}
\end{figure}

The HI 21cm signal from the IGM during reionization can be predicted
both analytically, using a standard Press-Schechter type of analysis
of linear structure formation, plus some recipes to approximate
non-linear evolution \cite{SCK05, ZFH04, Ali05, BA04, GS03}, or
through the use of numerical simulations \cite{CM03, FSH03}.

Fig 3 shows the expected evolution of the HI 21cm signal during
reionization based on the numerical simulations of \cite{FSH03}. They
find that the mean HI signal is about $T_B \sim 25$ mK prior to
reionization, with fluctuations of only a few mK on arcmin scales due
to linear density evolution.  In this simulation, the HII regions
caused by galaxy formation during reionization are seen in the
redshift range $z \sim 8$ to 10, reaching scales up to 2$'$
(frequency widths $\sim 0.3$ MHz $=> 0.5$ Mpc physical size). These
regions have (negative) brightness temperatures up to 20 mK relative
to the mean HI signal.  This corresponds to 5$\mu$Jy beam$^{-1}$ in a
2$'$ beam at 140 MHz.

We compare this signal to the sensitivity of a radio array.
The point source rms sensitivity (dual polarization) 
in an image from a synthesis radio telescope  is given by: 
\begin{equation}
{\rm rms} = ~
(\frac{70}{(\Delta\nu_{kHz} t_{hr})^{0.5}}) (\frac{T_{sys}}{300 {\rm K}})
(\frac{0.60}{\epsilon_{eff}}) (\frac{500 {\rm m}^2}{A_{ant}})
(\frac{27}{N_{ant}}) ~~ \rm{mJy ~ beam^{-1}}
\end{equation}
\noindent where $\Delta \nu$ is the channel width in kHz, $t$ is the
integration time in hours, $T_{sys}$ is the total system temperature,
$A$ is the collecting area of each element in the array, $\epsilon$ is
the aperture efficiency, and $N$ is the number of elements ($\epsilon
A N$ = total effective collecting area of the array). At low frequency
the value of $T_{sys}$ is dominated by the non-thermal foreground, and
again behaves as $T_{sys} \sim 100 (\frac{\nu}{200 \rm{MHz}})^{-2.8}$ K.
The beam FWHM is given by the Fourier transform of the uv 
coverage, assuming equal weight for each visibility. 

Consider an array with an effective collecting area of 1 square
kilometer at 140 MHz, distributed over 4 km, a system temperature of
300 K, a channel width of 0.3 MHz, and integrating for one month. The
rms sensitivity is then 1.3$\mu$Jy beam$^{-1}$, with a beam FWHM $\sim
2'$. This square kilometer array will be just adequate to perform true
three dimensional imaging of the average structure of the IGM during
reionization.

Unfortunately, the nearer term low frequency 'path finder' arrays will
have $\le 10\%$ the collecting of the SKA, and will likely not be able
to perform such direct 3D imaging (Table 1).  However, these near term
experiments should have enough sensitivity to perform power spectral
analyses of the HI 21cm fluctuations.  For power spectral analyses the
sensitivity is greatly enhanced relative to direct imaging due to the
fact that the universe is isotropic, and hence one can average the
measurements in annuli in the uv-plane, ie. the statistics of
fluctuations along a uv annulus are equivalent. The width of
independent uv annuli is set by the primary beam diameter, D, or
$\Delta l = 2 \pi \frac{D}{\lambda_{obs}}$. The situation is analogous
to COBE and WMAP. COBE lacked the sensitivity to make a true image
of the CMB fluctuations, but was able to make a robust determination
of the power spectrum of fluctuations.  WMAP was able to produce a
proper image of the CMB.

Many authors have considered the power spectrum of brightness temperature
fluctuations in the HI 21cm line
\cite{SCK05, ZFH04, Ali05, BA04, Morales04, SKES98, BMH05}.
It can be shown that the noise power spectrum, in mK$^2$, 
in standard spherical harmonic units,
$\frac{l(l+1)}{2 \pi} C_l^N$, for an array with uniform
coverage of the uv plane, is given by:

\begin{equation}
C_l^N = \frac{T_{sys}^2 (2\pi)^3}{\Delta \nu t f_c^2 l_{max}^2}
\end{equation}

\noindent where $f_c$ = areal covering factor of the array $= N_{ant}
\frac{A_{ant}}{A_{tot}}$, and $A_{tot}$ is the total area of the area
defined by the longest baseline. For example, for the logarithmic
antenna spacing of the VLA in the smallest configuration, the VLA-VHF
system has $f_c = 0.017$ on baselines out to 1 km, and $f_c = 0.17$
out to 0.25 km, for a full synthesis observation.  For comparison, the
SKA will have 30$\%$ of it's collecting area inside 1 km, or $f_c =
0.3$ out to 1 km. 

Most predictions of the HI 21cm power spectrum during reionization
published thus far parallel the 2D calculations applied to the CMB.
Ultimately, higher sensitivity, and physical insight, will come
through full 3D analysis of the HI 21cm fluctuations during
reionization \cite{Morales04}.  Of course, at the smallest $l$,
approaching the primary beam size $\Delta l$, the signal-to-noise is
limited by cosmic variance to essentially the number of synthesized
beams per primary beam.

\begin{figure}[!t]
\centerline{\psfig{file=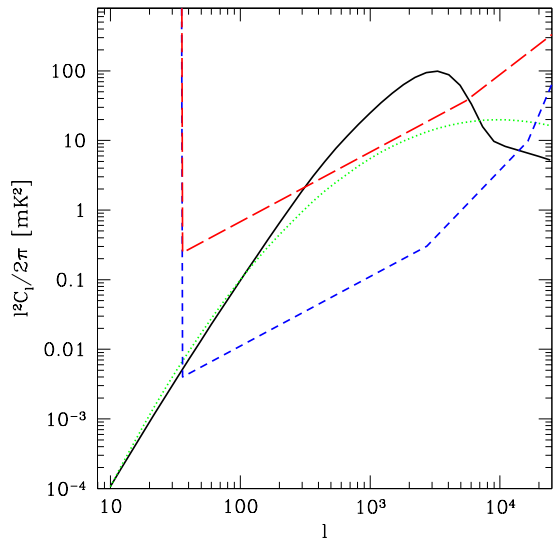,width=3in}}
\caption{Predicted HI 21cm brightness temperature power spectrum 
during reionization at $z=10$ \cite{FB04, ZFH04}. 
The green dotted line is the model prediction
including only linear density fluctuations. The black line solid
includes the effect of HII regions during reionization. The red
long dash line shows the sensitivity of pathfinder experiments like
LOFAR, while the blue short dash curve shows the sensitivity for the
SKA, assuming log $l$ bins. The cutoff at low $l$ is set by the 
primary beam.}
\end{figure}

As an example, Fig 4 shows the predicted evolution of the HI 21cm
power spectrum using an analytic calculation of structure formation,
including a model of (uncorrelated) HII region formation during
reionization \cite{FSH03, ZFH04} (see also \cite{BA04, SCK05, Wang05}
for analytic solutions, and \cite{FSH03, CM03} for predictions based
on numerical simulations). Fig 4 includes the power spectrum arising
from linear density fluctuations following the dark matter, plus the
effect of reionization.  The signature of reionization can be seen as
a bump in the power spectrum above the density-only curve due to the
formation of HII regions.  In this calculation, the rms fluctuations
at $z = 10$ peak at about 10 mK rms (observing frequency = 130 MHz) on
scales $l \sim 3000$\footnote{or $u = \frac{l}{2\pi} = 500$, or
$\theta \sim \frac{180^o}{l} = 3.6'$, or baselines, $b = \lambda u = 1.2$
km, or wavenumber $k \sim 
\frac{l}{10^4} {\rm Mpc}^{-1} = 0.3 {\rm Mpc}^{-1}$
(comoving).}. Note that in this calculation the neutral fraction as a
function of redshift is a free parameter, and is assumed to be 0.5 at
$z=10$. \cite{BL04} point out that there is no very large scale
information contained in the HI 21cm fluctuations, and that power
spectral analyses are best done on scales ranging from $l = 100$ to a
few thousand, ie. on angular scales $\le$ few degrees.

Also included are the noise power spectra for near and longer term
radio arrays. Typical noise values for near-term arrays are $\sim 1$
to 10 mK rms in the range $l = 10^3$ to $10^4$, for long integrations.
This should be adequate to determine the HI 21cm power spectrum during
reionization. The SKA will go an order of magnitude deeper, providing
an accurate measure of the HI power spectrum, and its evolution with
redshift. 

\cite{BL05, Ali05} have shown that gas peculiar velocities, ie.
infall into superclusters and along filaments (analogous to the Kaiser
effect for galaxy clustering), increases the expected rms fluctuations
by about a factor two.  Also, clustering of luminous sources
(ie. biased galaxy formation) implies a characteristic scale for the
HII regions of 1 to 10 Mpc (comoving), and leads to another factor two
in fluctuation strength \cite{FMH05, SCK05} over the random
distribution of HII regions assumed by \cite{ZFL04}. Specifically,
this later effect moves the fluctuation power to larger scales.

\cite{ZFL04, BA04, FSH03} consider the effect of spectral channel
width on the power spectral analysis.  Bandwidth enters via the finite
thickness of the HII regions and the density fluctuations.
They conclude that a bandwidth of about 0.2 MHz is optimal for power
spectral studies, although the rms of the signal changes by only a
factor of about two going from 0.1 to 0.5 MHz channel width.

\cite{FSH03} show how the power spectra change with different
reionization models, including outside-in versus inside-out models,
uniform versus HII region dominated reionization, and double
reionization. They point out that HI 21cm power spectral studies
should provide powerful constraints on the process of reionization,
with different reionization scenarios clearly discernable in their
power spectral signature. Note that this is unlike CMB large scale
polarization measurements, which provide reasonable constraints on the
total Thompson scattering optical depth through the universe back to
recombination, but only marginal constraints on the reionization
history \cite{HuHolder03}.

Lastly, \cite{BA04, nusser04} point out that if full 3D tomography of the
evolution of IGM structures could be made, then one might also
constrain the geometry of universe through the standard
Alcock-Pacinsky effect, ie. the difference in structure evolution
observed in angle versus along the line-of-sight, due to the
non-Euclidean geometry of the univese. 

\subsubsection{Clustering of minihalos}

A point of debate has been the fraction of HI in collapsed objects, as
opposed to the diffuse IGM \cite{Iliev03}. This fraction has a complex
depence on structure formation history, and \cite{OM03, Gnedin04}
conclude that the majority of the HI during reionization will remain
in the diffuse phase.  However, at times when $T_S$ approaches
$T_{CMB}$ in the diffuse IGM, one might expect HI 21cm emission due to
clustering of minihalos\footnote{Minihalos have masses
$< 10^7$ M$_\odot$, and virial temperatures $< 10^4$ K.  These halos
cannot cool via atomic hydrogen lines, and therefore 'saturate' at
over-densities of $\delta \sim 100$. They will not form stars unless an
alternate cooling mechanism can be found. One candidate is molecular
hydrogen, which is an effective coolant down to virial temperatures of
100 K \cite{BL00}.}, in which $T_S > T_{CMB}$ simply due to the
virialization process \cite{FSH03}.

\cite{Iliev03} have considered the HI 21cm power spectrum due to
clustering of minihalos at high redshift.  For a beam size of 1$'$,
and channel width of 0.2 MHz, they predict 3$\sigma$ brightness
temperature fluctuations due to clustering of minihalos $\sim 7$mK at
$z = 8.5$, decreasing to 2mK at $z=20$.

\subsubsection{Other tests}

\cite{CF05} discuss circular polarization of the 21cm line due to
zeeman splitting during reionization. Such polarization would be
detectable with the SKA if the IGM field strength were 100 $\mu$G.
\cite{Salvaterra05, Cooray04} consider the anti-correlation between HI
21cm fluctuations and sub-degree scale CMB secondary anisotropies due
to Thompson scattering in ionized regions during reionization at $l >
1000$ (kinetic SZ effect).  A natural anticorrelation would be
expected between the neutral and ionized regions. And lastly,
\cite{SC05} discuss delensing of the CMB using the HI 21cm
fluctuations to pin-point density inhomogenaeities along the line of
sight ($z \sim 10$ to 200).  This could prove important for decoupling
the E and B mode mixing that occurs due to lensing, possibly
recovering the intrinsic (inflationary) B polarization.

\subsubsection{Cosmic Stromgren spheres}

While direct detection of the typical structure of HI and
HII regions may be out of reach of the near-term 21cm telescopes,
there is a chance that even this first generation of
telescopes will be able to detect the rare, very large
scale HII regions associated with luminous quasars
near the end of reionization.

The presence of cosmic Stromgren spheres around the highest redshift
SDSS QSOs has been deduced from the observed difference between the
redshift of the onset of the GP effect and the systemic redshift of
the host galaxy \cite{White03, WL04a, WLC05, Walter03} (although
cf. \cite{OF04}). The physical size of these spheres is $\sim 5$ Mpc,
or an order of magnitude larger than the typical spheres expected from
clustered galaxy formation, due to the extreme luminosity of the
sources ($\sim 10^{14}$ L$_\odot$). If the neutral fraction remains
substantial to relatively low redshift, f(HI) $> 0.1$ \cite{MH03,
WL04a, WLC05}, then it is plausible to search for these regions as
'holes in the sky', or regions of negative brightness in the redshift
21cm line frequency.  The expected signal is $\sim 20$mK $\times$ f(HI)
on a scale $\sim 10'$ to 15$'$, with a line width of $\sim 1$ to 2 MHz
\cite{WL04b}. This corresponds to 0.5 $\times$ f(HI) mJy beam$^{-1}$,
for a 15$'$ beam.

Searching for HI signals around known high redshift QSOs has a number
of major observational advantages: (i) the exact location in frequecy,
and RA and DEC, are well determined, thereby limiting the search space
dramatically, and (ii) there is already evidence for the existence of
the features based on the Ly$\alpha$ spectra.  The VLA-VHF system has
been specifically design to search for these spheres around known SDSS
quasars at $z \sim 6$ to 6.4 \cite{VHF}. Fig 5 shows the simulated
image and spectrum of such a Stromgren sphere around an SDSS QSO
assuming f(HI) = 1, as observed with the VLA-VHF system.  While it is
unlikely the mean neutral fraction is this high at such a low
redshift, in 250 hours the VLA-VHF system should set the first direct
constraints on neutral fraction of the IGM, at the level f(HI) $\ge
0.2$, as well as easily rule out more extreme models, such as $T_S <<
T_{CMB}$.  \cite{WLB05} present similar simulations for the MWA and
the SKA.

\begin{figure}[!t]
\psfig{file=VLAD-EOR-250.PS,width=2.2in}
\vskip -2.4in
\hspace*{2.4in}
\psfig{file=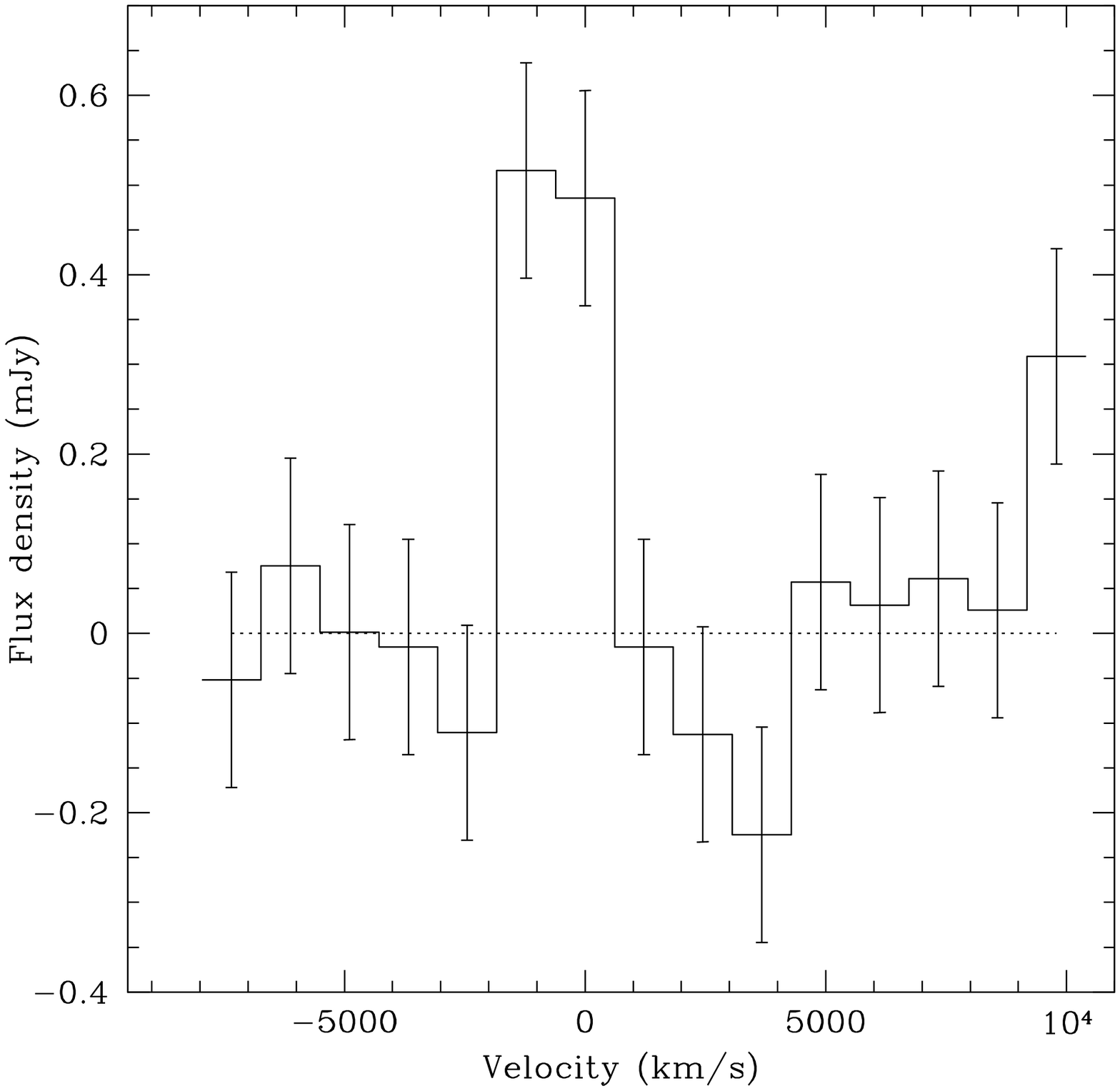,width=2.2in}
\caption{
Simulated image and spectrum
of a Stromgren sphere surrounding
a luminous SDSS QSO at the end of reionization 
for a 250 hour integration with the VLA-VHF system \cite{VHF}. 
The contour levels are: -0.45, -0.3, -0.15, 0.15, 0.3, 0.45 mJy beam$^{-1}$,
and the negative signal has been inverted to appear positive. 
}
\end{figure}

\cite{Kohler05} also discuss spectral dips due to large HII regions
around luminous quasars ($> 2\times 10^{10}$ L$_\odot$) during the
EoR.  Their Fig 2 shows a spectrum in a typical 10$'$ beam from 100 to
180 MHz. They predict on average one relatively deep (-2 to -4 mK) dip
per LoS on this scale. They emphasize that these spectral features may
be easier to detect than spatial fluctuations of a similar magnitude,
due to the fact that spatial confusion is highly structured on arcsec
to arcmin scales, while the spectral confusion should be smooth over
10's of MHz.

\cite{WLB05} perform a similar calculation using the evolution of the
bright QSO luminosity function to predict the number of HII regions
around active QSOs at $z > 6$. They conclude that there should be
roughly one SDSS-type HII region around an active QSO (physical radius
$>$ 4Mpc) per 400 deg$^2$ field per 16 Mhz bandwidth at $z \sim 6$,
and one $R \ge 2$ Mpc region at $z \sim 8$. They also point out that
the recombination time may be longer than the Hubble time, so that
fossil HII regions may be observed around non-active AGN. Assuming a
duty cycle $\sim \frac{<t_{qso}>}{t_{H}(z)} \sim \frac{10^7}{10^9}
\sim 0.01$, where $<t_{qso}>$ is the fiducial lifetime of QSO activity
\cite{YT99}, and $t_{H}(z)$ is the age of the universe, leads to a
factor 100 more fossil HII regions per FoV.

\subsubsection{Beyond reionization}

A number of authors have considered the brightness temperature
fluctuations due to the 21cm line from the neutral IGM
at redshifts prior to reionization, $z > 20$.
\cite{BL04, BL05} predict the power spectrum of fluctuations during
the era when the Ly$\alpha$ photons from the first luminous objects
couple $T_S$ to $T_K$ locally, but the universe remains substantially
neutral ($z \sim 20$ to 30), and $T_K < T_{CMB}$ in the diffuse IGM.
Brightness temperature fluctuations can be due to emission from
clustered minihalos, plus enhanced by absorption against the CMB by
the diffuse IGM, with an rms $\sim$ 10mK for $l \sim 10^5$, due to a
combination of linear density fluctuations, plus poisson ('shot')
noise and biasing in the Ly$\alpha$ source (ie. galaxy) distribution.

\cite{LZ04} go even further in redshift, to $z > 50$ to 200. In this
regime the HI generally follows linear density fluctuations, and hence
the experiments are as 'clean' as CMB studies, and $T_K < T_{CMB}$, so
a relatively strong absorption signal might be expected.  \cite{LZ04}
also point out that Silk damping, or photon diffusion, erases
structures on scales $l > 2000$ in the CMB at recombination,
corresponding to comoving scales = 22 Mpc. The HI 21cm measurements
can explore this physical regime at $z \sim 50$ to 300. The predicted
rms fluctuations are 1 to 10 mK on scales $l= 10^3$ to 10$^6$ (0.2$^o$
to 1$''$).  These authors point out that, due to sensitivity to large
$l$, and the 3D nature of the information, HI 21cm power spectral
studies in this epoch ``contain an amount of information orders of
magnitude larger than any other cosmological probe''. These data could
provide the best tests of non-Gaussianity of density fluctuations, and
for constrainting the tilt, or running power law index, of mass
fluctutions to large $l$, providing important tests of inflationary
structure formation. \cite{sethi05} also suggests that a large global
signal, up to -0.1 K, might be expected for this redshift range.

Two experiments are being planned to cover this low frequency range,
the long wavelength array (LWA) in New Mexico, and in the longer term,
LUDAR is being considered for the back side of the moon.

\subsection{Absorption toward discrete radio sources}

Observing HI 21cm emission from the EoR implies studying large
scale structure (cluster scales and larger).  A number of groups
have recently considered the possibility of studying smaller scale
structure in the neutral IGM by looking for HI 21cm absorption toward
the first radio-loud objects (AGN, star forming galaxies, GRBs)
\cite{Carilli04a}.

\cite{CGO02} use numerical simulations to predict
the HI 21cm absorption profile of the 'cosmic web' prior to
reionization. For example, for a source at $z = 10$, they predict an
average optical depth due to 21cm absorption of about 1$\%$,
corresponding to the 'radio Gunn-Peterson effect'. They also find
about five narrow (few km/s) absorption lines per MHz with optical
depths of a few to 10$\%$ (Fig 2b). These latter lines are
equivalent to the Ly $\alpha$ forest seen after reionization, and
correspond to over-densities evolving in the linear regime ($\delta
\le 10$).  While significant questions remain about simulating the
thermal state of the IGM to such detail, the simple point remains
that, while the Ly$\alpha$ lines from such structures are highly
saturated, the (much) lower 21cm oscillator strength makes the IGM
translucent prior to reionization.

\cite{FL02, OH03} predict a similar HI 21cm absorption line density due
to gas in minihalos as that expected for the 21cm forest.  
\cite{Cen02} shows that the presence or absence of HI absorption by
minihalos, or the cosmic web, would be a telling diagnostic of early
IGM heating mechanisms.
\cite{FL02} also consider the expected 21cm
absorption profiles for proto-disk galaxies. While such absorption
lines will be rare ($10^4$ times less frequent than the 21cm forest
lines), the optical depths may be large enough that the lines could be
observed toward faint radio sources, in particular, gamma ray burst
radio after-glows within the host galaxy \cite{IM05}.

An important caveat in these calculations is the assumption of radio
loud sources during the EoR.  This question has been considered in
detail by \cite{CGO02, haiman04, JR05}.  They show that current
models of radio-loud AGN evolution predict between 0.05 and 1 radio
sources per square degree at $z > 6$ with $\rm S_{150MHz} \ge 6$ mJy,
adequate for EoR HI 21cm absorption studies with the SKA.

\section{Observational challenges}

\subsection{Foregrounds}

It has long been recognized that the HI 21cm signal from reionization
must be detected on top of a much larger non-thermal (synchrotron)
signal from foreground emission. This foreground includes discrete
radio galaxies, and large scale emission from our own Galaxy. The
expected HI signal is about $10^{-4}$ of the foreground emission.

Many groups have considered the effects of the foregrounds on HI 21cm
EoR imaging and power spectral studies.  \cite{dMat01} show that, even
if point sources can be removed to the level of 1 $\mu$Jy, the rms
fluctutions on spatial scales $\le 10'$ ($l \ge 1000$) due to residual
radio point sources will be $\ge 10$ mK just do to Poisson
noise, increasing by a factor 100 if the sources are strongly
clustered. Clearly this conclusion depends on the
extrapolation of radio source populations to $\le 1\mu$Jy, but under
reasonable assumptions, even in the Poisson case the residual source
fluctuations will be comparable to the HI signal.

This calculation has led many groups to consider removal of
foregrounds in the spectral domain. The important point is that the
foregrounds should be relatively smooth in frequency,
predominantly the sum of power-law, low frequency non-thermal spectra,
or perhaps gently curving on spectral scales $\ge 10$'s MHz. The HI
signal should show significant structure on sub-MHz scales,
corresponding to the typical size scale for features during
reionization.

A number of complimentary approaches have been presented for
foreground removal.  \cite{GS03, Wang05} consider removal of the
foregrounds through fitting of smooth spectral models (power laws or
low order polynomials in log space) to the observed visibilities or
images. \cite{MH03, Morales04} present a 3D Fourier analysis of the
measured visibilities, where the third dimension is frequency. They
show that the different symmetries in this 3D space for the signal
arising from the noise-like HI emission, versus the smooth (in
frequency) foreground emission, can be a powerful means of
differentiating between foreground emission and the EoR line signal.
\cite{SCK05, BA04, ZFH04} perform a similar analysis, only in the
complementary Fourier space, meaning cross correlation of spectral
channels. They show that the 21cm signal will effectively decorrelate
for channel separations $> 1$ MHz, while the foregrounds do not, and
this too can be a means of separating the foregrounds from the desired
21cm fluctuations. The overall conclusion of these methods is that
spectral decomposition should be adequate to separate non-thermal
foregrounds from the HI 21cm signal from reionization at the mK level.

However, there are two potentially more insideous observational
challenges relating to non-thermal foregrounds beyond simple in-beam
confusion.  First, there can be frequency dependent sidelobes from
confusing sources in the very wide fields being considered. A
particularly problematic effect has been the recent observation that
telescope spectral response (ie. bandpass) may depend on position of a
given source in the primary beam \cite{Oosterloo05}. The origin of
this effect remains uncertain (perhaps relating to scattering and
blocking structures), but it will lead to frequency dependent
sidelobes that potentially affect widefields and are not removed
through normal (on-axis) bandpass calibration, and may require a
spatially dependent bandpass calibration.

And second is polarized structure in the diffuse emission in the
field.  Wide field polarization has been seen with low frequency
observations with the WSRT at 330 MHz and below, and is thought to
arise due to differential Faraday depth through the ISM
\cite{haverkorn}. Hence, great care is required to control,
and calibrate, the wide field polarization response of the primary
elements.

\subsection{Ionosphere}

A second potential challenge to low frequency imaging over wide fields
is phase fluctuations caused by the ionosphere.  These fluctuations
are due to index of refraction fluctuations in the ionized plasma, and
behave as $\Delta \phi \propto \nu^{-2}$.  Morever, the typical
'isoplanatic patch', or angle over which a single phase error applies,
is a few to 10 degrees (physical scales of 10's km in the ionosphere),
depending on frequency \cite{Cotton04, lane}.  Fields larger than the
isoplanatic patch will have multiple phase errors across the field,
and hence cannot be corrected through standard (ie. single solution)
phase self-calibration techniques.

New wide field self-calibration techniques, involving multiple phase
solutions over the field, or a 'rubber screen' phase model
\cite{Cotton04, hopkins}, are being developed that should allow for
self-calibration over wide fields. However, even self-calibration
techniques will be insufficient to overcome the very rapid
variations caused by the occasional ionospheric storm, or 
traveling ionospheric disturbances.

At very low frequency ($\le$ a few MHz), one approaches the plasma
frequency of the ionosphere, and the optical depth increases
dramatically, precluding observations from the ground.

\subsection{Interference}

Perhaps the most difficult problem facing low frequency radio
astronomy is terrestrial (man-made) interference. The relevant
frequency range corresponds to 7 to 200 MHz ($z= 200$ to 6). These are
not protected frequency bands, and commercial allocations include
everything from broadcast radio and television, to fixed and mobile
communications.

Many groups are pursuing methods for RFI mitigation and excision (see
extensive references at \cite{RFI}). These include: (i) using a
reference horn, or one beam of a phased array, for constant monitoring
of known, strong, RFI signals, (ii) conversely, arranging
interferometric phases to produce a null at the position of the RFI
source, and (iii) real-time RFI excision using advanced filtering
techniques in time and frequency, of digitized signals both pre- and
post-correlation. This requires very high dynamic range (many bit
sampling), and very high frequency resolution.

One obvious advantage of HI absorption experiments (section 3.3) is
that they can be done using long baselines (10's to 100's of km).
Long baselines lead to decorrelation of the terrestrial signal due to
fringe tracking of the celestial source.  However, the fringe rates on
the short baselines required for the HI emission experiments ($\le$
few km) are low, such that decorrelation due to fringe winding will
not be a very effective RFI filter on the shorter baseline.

In the end, the most effective means of reducing interference is to go
to the remotest sites. \cite{MWA, PAST} have selected sites in
remote regions of Western Australia, and China, respectively,
because of known low RFI environments. Of course, the
ultimate location would be the back side of the moon.

\section{Telescopes}

Many programs have been initiated to study the HI 21cm signal from
cosmic reionization, and beyond. These are summarized in Table 1. A
number of different approaches are being taken, both techinically and
in order to address different aspects of the problem \cite{Morales04}.

The largest near-term efforts are the Mileura Wide Field Array (MWA),
the Primeval Structure Telescope (PAST), and the Low Frequency
Array (LOFAR). These are being optimized to study the power spectrum
of the HI 21cm fluctuations, although in principle they will be able
to image the larger HII regions during reionization. The VLA-VHF
system is designed specifically to set limits on the HII regions
around $z\sim 6$ to 6.4 SDSS QSOs, although it should also constrain
the late-time power spectrum.  In the long term the Square Kilometer
Array should have the sensitivity to perform true three dimensional
imaging of the neutral IGM in the 21cm line during reionization.
And at the lowest frequencies ($< 50$ MHz), 
the Long Wavelength Array (LWA), and eventually the lunar array
(LUDAR \cite{LUDAR, maccone}), are being designed for the higher 
$z$ signal, prior to reionization.

As discussed above, the technical challenges are many. Use of spectral
decomposition to remove the foregrounds requires careful control of
the synthesize beam as a function of frequency, with the optimal
(although difficult) solution being a telescope design where the
synthesized beam is invariant as a function of frequency \cite{MarkI}.
High dynamic range front ends are required to avoid saturation in
cases of strong interference, while fine spectral sampling is required
to avoid Gibbs ringing in the spectral response.  The polarization
response must be stable and well calibrated. Calibation in the
presence of a significant ionospheric phase screen requires new wide
field calibration techniques.  The very high data rate expected for
many element ($\ge 10^3$) arrays requires new methods for data
transmission, cross correlation, and storage \cite{MWA}.  At the
lowest frequencies, $\le 20$MHz or so, where we hope to study the
pre-reionization IGM, phase fluctuations and the opacity of the
ionosphere becomes problematic, leading to the proposed LUDAR project
on the far side of the moon. The far side of the moon is also the best
location in order to completely avoid terrestrial interference.

\begin{table}
\scriptsize
\caption{HI 21cm Experiments \label{tab:Experiments}}
\begin{center}
\begin{tabular}{cccccccccc}
\noalign{\hrule}
~ & Freq & Area & B$_{max}$ & Site & Type & FoV & Date & Goal & Ref \\
~ & MHz & 10$^4$ m$^2$ & km & ~ & ~ & deg & ~ & ~ & ~ \\
\noalign{\hrule\hrule}
GMRT & 150 -- 165 & 3.7 & 10 & India & Parabola & 4 & 2000 & StromSph &
\cite{GMRT} \\
PAST & 50 -- 200 & 7 & 2 & China & Dipole & 10 & 2006 & PowSpec &
\cite{PAST} \\
VLA-VHF & 180 -- 200 & 1.3 & 1 & USA & Parabola & 4 & 2006 & StromSph &
\cite{VHF} \\
Mark I & 100 -- 200 & ~ & 0 & Aus & Spiral & 180 & 2006 & Global
& \cite{MarkI} \\
MWA-LFD$^1$ & 80 -- 300 & 3 & 1.5 & Aus & Dipole & 20 & 2007 & PowSpec &
\cite{MWA} \\
LOFAR$^2$ & 115 -- 240 & 10 & 2km=40\% & NL & Tiles & 20 & 2007 & PowSpec
& \cite{LOFAR} \\
LWA & 10 -- 88 & 10 &  5km=30\% & USA & Dipole & 20 & 2008 &
Pre-reion & \cite{LWA} \\
SKA & 100 -- 200 & 100 & 5km=50\% & ? & Dipole & 20 & 2015 & Imaging &
\cite{SKA} \\
LUDAR & 5 -- 50 & 1000 & 100 & Moon & Dipole & 60 & 2020 & pre-reion &
\cite{LUDAR} \\
\noalign{\hrule}
\end{tabular}

\end{center}

\noindent ~$^1$An order of magnitude increase in area is
planned after testing the Low Frequency Demonstrator.

\noindent ~$^2$LOFAR will also have a low frequency component between
30 and 80 MHz.
\end{table}

Acknowledgements: Thanks to the Max-Planck Society and
the Alexander von Humboldt Foundation for support through
the Max-Planck Research Prize, and S. Furlanetto and N. Gnedin for 
permission to use figures.

\end{document}